\documentstyle [nato,numreferences,epsf]{crckapb}

\global\arraycolsep=2pt%
\def\lsim {\,\,\vcenter{\hbox{$\buildrel{\displaystyle\lt}\over\sim$}}\,\,}
\def\gsim {\,\,\vcenter{\hbox{$\buildrel{\displaystyle\gt}\over\sim$}}\,\,}
\def\coeff	#1#2{{\textstyle{#1\over #2}}}%
\def\half	{\coeff {1}{2}}%
\def\sym	{_{\rm sym}}
\def\asym	{_{\rm asym}}
\def\eps	{\epsilon}
\def\mh		{M_{\rm H}}
\def\mw		{M_{\rm W}}
\def\Tc		{T_{\rm c}}
\let\citenum=\cite

\mathchardef\lt="313C%
\mathchardef\gt="313E%
\mathcode`\@="8000 {\catcode`\@=\active \gdef@{{\dagger}}}%
\mathcode`\<="8000 {\catcode`\<=\active \gdef<{{\delimiter"426830A}}}%
\mathcode`\>="8000 {\catcode`\>=\active \gdef>{{\delimiter"526930B}}}%

\begin {opening}
\begin {title}
    {%
    The Electroweak Phase Transition:
    \protect\\A Status Report%
    $^*$
    }%
\end {title}

\author
    {%
    Laurence G. Yaffe$^*$
    }
\institute
    {%
    University of Washington
    \\Department of Physics
    \\Seattle, Washington 98195-1560 USA
    }
\end {opening}

\runningtitle {The Electroweak Phase Transition}

\begin {document}

\setcounter{footnote}{1}
\footnotetext
    {%
    Presented at {\sl String Gravity and Physics at the
    Planck Energy Scale}, Erice, September 1995.
    }
\setcounter{footnote}{2}
\footnotetext
    {%
    Research supported in part by Department of Energy
    grant DE-FG06-91ER40614.
    }
\setcounter{footnote}{0}

    The possibility that the observed cosmological baryon density
might be a consequence of non-perturbative dynamics at the
electroweak phase transition is one of the most exciting
topics at the interface of particle physics and cosmology.
However, determining the viability of electroweak baryogenesis
in (minimal extensions of) the standard model requires knowledge of
the equilibrium behavior of the electroweak phase transition,
non-equilibrium dynamics around expanding bubble walls,
and non-perturbative baryon violating processes in both
the high and low temperature phases~\cite {baryogenesis}.

    This talk will examine the current understanding of the
electroweak phase transition.
Determining the order of the phase transition, and the magnitudes
of the latent heat, correlation lengths, and the baryon violation
rate at the transition are some of the key questions.
Viable baryogenesis scenarios require a first order transition
with rapid suppression of baryon violating transitions
in the low temperature phase \cite {baryogenesis,Dine,Shaposhnikov}.

    Performing reliable quantitative calculations of electroweak
transition properties is challenging,
in part because there is important dynamics, both perturbative
and non-perturbative, on many differing length scales.
Approximation methods which have been applied to this problem
include weak coupling perturbation theory (or mean field theory)
\cite {Dine,Arnold&Espinosa,Bagnasco&Dine},
$\epsilon$-expansions \cite {A&Y,Alford,Gleisser&Kolb},
and numerical simulations \cite {Kripfganz,Kajantie,Farakos,Fodor&Co}.
Each of these approaches have significant (but differing) limitations;
for example, ordinary perturbation theory is valid if
the Higgs mass is much lighter than the $W$-boson mass,
but does not appear to be trustworthy for the experimentally allowed
range of possible Higgs masses.
Nevertheless, all of these approaches have helped to elucidate
the rich interplay of the physics at the electroweak phase transition.

    I will focus on the behavior of the minimal standard model,
despite the fact that electroweak baryogenesis in the strictly minimal
model cannot explain the observed cosmological baryon density.
There are two basic problems with baryogenesis in the minimal model.
For experimentally allowed values of the Higgs mass,
most of the baryon excess which is produced at the phase transition is
destroyed shortly thereafter \cite {Dine}.
And even if that weren't the case,
the baryon excess produced in the minimal
standard model would still be far too small to agree with present day
observations~\cite {baryogenesis}.
The first problem will be discussed further below.
The second problem is a consequence of the nature of CP violation in
the minimal model.
Without fundamental CP violation, no baryon excess can be dynamically
produced.
But in the minimal standard model, CP violation arises only through
a phase in the CKM matrix elements, and can only appear in processes
involving all three generations of fermions.
Because of the freedom to redefine the relative phases of fermion fields,
one may show that any CP violating transition must involve a
reparameterization invariant combination of Yukawa couplings and
mixing angles which is tiny \cite {Shaposhnikov},
\begin {equation}
    \delta_{\rm CP} \lsim 10^{-20} \,.
\end {equation}
Given this suppression, simple estimates show that it is essentially
impossible to produce a baryon-to-photon ratio of the required $10^{-10}$
magnitude~\cite {baryogenesis}.

Both of these difficulties can be avoided in extensions
of the minimal standard model which incorporate,
for example, a single additional scalar field.
This allows explicit CP violation in the scalar sector,
and can increase the sphaleron mass and thereby
slow down the rate of baryon-violating reactions
just after the transition.
However, the challenges involved in analyzing the electroweak phase transition
are, for the most part, generic to any electroweak theory.
Consequently, the minimal standard model provides a useful toy model
(with the fewest adjustable parameters) which may be used as a testing
ground for quantitative calculations of electroweak
phase transition properties.

\section {Perturbation Theory}

    Since the standard model is weakly coupled (at scales of several
hundred GeV), perturbation theory is the most obvious approach for studying
the electroweak phase transition.
A one-loop calculation of the effective potential for the Higgs field
is elementary and gives (schematically)
\begin {eqnarray}
    V(\phi) &=& V_{\rm tree}(\phi)
    + T \int d^3k \> \ln \left(1 - e^{-E(k)/T} \right)
\label {one-loop-form}
\\ &=&
    (-\mu^2 + a \, g^2 T^2) \, \phi^2 - b \, g^3 T \, \phi^3
		+ \lambda \phi^4 + \hbox {($\phi$-independent)}
\label {one-loop-pot}
\end {eqnarray}
For simplicity, only the thermal $W$-boson contribution is indicated in
(\ref {one-loop-form}).
The $W$-boson energy $E(k) \equiv \sqrt {k^2 + \mw(\phi)^2}$ with
$\mw(\phi) \equiv \half g \phi$,
and $a$ and $b$ are dimensionless
constants which I will henceforth suppress.

\begin {figure}[h]
\vbox
    {%
    \begin {center}
	\leavevmode
	
	\epsfbox [150 260 500 530] {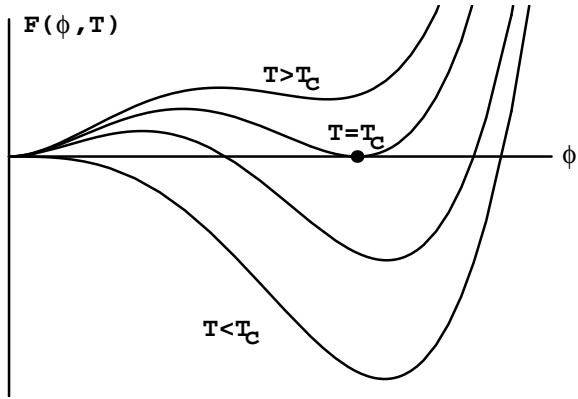}
    \end {center}
    \caption
	{%
	\label {Vfig}
        The form of the free energy, as a function of $\phi$, for
        different temperatures.
	}%
    }%
\end {figure}

The $O(T^2)$ thermal correction to the quadratic term in the potential
is responsible for driving the transition
from the Higgs phase, with $<\phi> \ne 0$,
to an unbroken symmetry phase with $<\phi> = 0$,
as the temperature is increased.
If corrections to this one-loop result are negligible,
then the presence of the $O(g^3 T)$ cubic term will cause the transition
to be first-order, with a barrier separating co-existing broken
and unbroken symmetry phases at the transition temperature $\Tc$.
(See Fig.~1.)\ %
In the Higgs phase near the transition,
the quadratic, cubic, and quartic terms
in the potential (\ref {one-loop-pot})
are all comparable in size.
Consequently, $\phi_{\rm c} \sim g^3 T / \lambda$, or
\begin {equation}
	 \left. {\mw \over g^2 T} \right|_{\Tc}
    \sim {g^2 \over \lambda}
    \sim \left. {\mw^2 \over \mh^2}  \right|_{T = 0} \,.
\label {one-loop-ratios}
\end {equation}

These ratios are important for several reasons.
First, the rate of non-perturbative baryon violating reactions
in the low temperature Higgs phase
is controlled by a thermal activation energy given by the
(temperature dependent) sphaleron mass,
\begin {equation}
    \Gamma_{\Delta {\rm B}} \sim T^4 \> e^{-M_{\rm sph}(T) / T} \,,
\end {equation}
and the sphaleron mass
$M_{\rm sph}(T)$ equals $16\pi \mw(T)/ g^2$ (to within a factor of 2).
Hence, the relation (\ref {one-loop-ratios}) implies that the
baryon violation rate, just after the transition, is exponentially
sensitive to the zero temperature ratio of Higgs and $W$-boson masses,
\begin {equation}
    \ln \Gamma_{\Delta {\rm B}} \sim
    - \left. {\mw^2 \over \mh^2} \right|_{T=0} \,.
\label {mass-bound}
\end {equation}
In order for electroweak baryogenesis to be viable,
baryon violating reactions after the transition must turn off
sufficiently rapidly so that the baryon asymmetry produced during
the transition can survive to the present day.
The simple result (\ref {mass-bound}) implies that this will only
be possible if the Higgs to $W$ mass ratio is sufficiently small.
More detailed analysis, based on the one-loop effective potential
in the minimal standard model,
produces an estimate of 35~Gev for the {\em upper} limit on the Higgs mass
if baryon violating rates after the transition are to be acceptably small
\cite {Dine}.
This, of course, is inconsistent with the current experimental {\em lower}
bound on the Higgs mass of about 65 GeV.

As noted above, one can circumvent this ``no-go'' result
if minor extensions (such as adding a second scalar field)
are made to the minimal standard model.
However, one should first ask whether the 35 GeV bound in the minimal model,
or any other conclusion based on the one-loop analysis
(including the predicted first order nature of the transition!)
is reliable.
To answer this, one must understand whether higher order corrections
to the one loop results are significant.

    Life would be simple if the temperature $T$ were the only relevant
scale for physics at the phase transition.
If this were the case, then higher order corrections would automatically be
suppressed by powers of $g^2$ (or $\alpha_{\rm W}$) at the scale $T$,
which is, in fact, small.%
\footnote
    {%
    This assumes that the Higgs is light enough to be weakly coupled.
    }
However, life is not so simple.
Consider, for example, any effective potential diagram
containing only gauge field lines.
The contribution of each loop may be estimated as%
\footnote
    {%
    Recall that Euclidean space frequencies are discrete
    at finite temperature, $q^0 = \omega_n \equiv (2 \pi n T)$.
    Convergent diagrams are dominated by the static $n {=} 0$
    frequency component.
    }
\begin {eqnarray}
    g^2 T \int d^3 q \> \left( {\bf q}^2 + \mw(T)^2 \right)^{-2}
    \sim
    g^2 T / \mw(T) \,.
\end {eqnarray}

Therefore the real loop expansion parameter is not $g^2$,
but rather equals $g^2 T / \mw(T)$.
Consequently, the reliability of perturbation theory,
in the Higgs phase near the transition,
is controlled by the the same ratio of
$$
    {g^2 \Tc \over \mw(\Tc)} \sim \left. {\mh^2 \over \mw^2} \right|_{T=0}
$$
which governs the baryon violation rate.
Perturbation theory is not reliable unless the physical Higgs
is sufficiently light.

    Of course, the above estimate does not determine whether
perturbation theory really breaks down at $\mh = \mw / 2$,
or at $\mh = 2 \mw$.
For a quantitative answer, one must actually compute higher
order corrections to various quantities of interest.
Fortunately, a number of two-loop calculations of phase
transition properties have now been performed.
Fig.~2 illustrates the result for the effective potential
at a Higgs mass of 35 GeV \cite {Arnold&Espinosa,Bagnasco&Dine}.
The magnitude of $<\phi>$ in the Higgs phase at the transition
shifts by only about 20\%, but the height of the barrier changes
by almost a factor of three!

\begin {figure}[h]
\vbox
    {%
    \begin {center}
	\leavevmode
	
	\epsfbox [150 250 500 550] {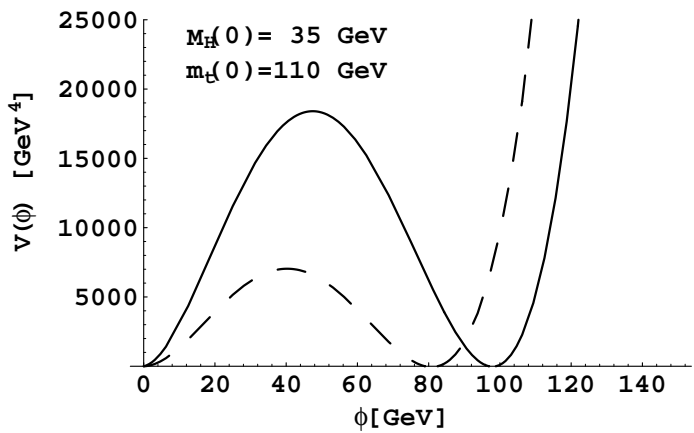}
    \end {center}
    \caption
	{%
	\label {figa}
	The effective potential at the critical temperature for
	$\mh(0)$ = 35 GeV and $m_{\rm t}(0)$ = 110 GeV.
	The dashed and solid lines
	are the one-loop and two-loop results respectively.
        [Why a 110 GeV top mass? Because this is an old graph.
        But the results aren't particularly sensitive to $m_{\rm t}$.]
	}%
    }%
\end {figure}

Since both the barrier height, and the Higgs expectation value,
are gauge dependent, one might legitimately wonder whether
truly physical quantities will be better behaved.
However, a two-loop evaluation of the dimensionless latent heat,
$\Delta Q / \Tc^4$, and surface tension, $\sigma / \Tc^3$,
at a Higgs mass of 50 GeV finds 50--100\% changes over the
one-loop results \cite {Fodor&Co}.
Consequently, there appears to be good reason to expect a Higgs mass
somewhere between 35 and 50 GeV to be the upper limit for the
reliability of perturbation theory for most quantities.%
\footnote
    {%
    Note, however, that the authors of ref.~\citenum {Farakos}
    claim that using the $3d$ renormalization group to improve
    some logarithmic corrections can delay the
    breakdown of perturbation theory.
    }
\noindent

Finally, even when the Higgs to $W$ mass ratio is small,
so that perturbation theory in the Higgs phase is well behaved,
the reliability of perturbative calculations of phase transition
properties is limited by the presence of non-perturbative physics
in the unbroken symmetry phase.
The long distance physics of the unbroken phase is described by a
three dimensional non-Abelian theory with a confinement scale
of $g^2 T$, and has a free energy density of order $(g^2 T)^3$ which is
incalculable in perturbation theory.
Consequently, for small values of the Higgs fields, where
$\mw(\phi) \lsim g^2 T$,
any perturbative approximation to the effective potential will have
a systematic uncertainty of order $g^6 T^4$.
This will generate an uncertainty in the value of the transition
temperature, or any other phase transition property.
However, when $\lambda / g^2$ is small,
this uncertainty is of the same order as the unknown
four-loop contributions to the free energy in the Higgs phase
\cite {Arnold&Espinosa}.

\section {Numerical Simulations}

    The electroweak phase transition can be also be studied
numerically by performing stochastic simulations in a lattice
version of the theory.
Simulating the full theory (with chiral fermions and SU(3) $\times$
SU(2) $\times$ U(1) gauge fields) is not practical.
However, since one is interested in high temperature physics,
all non-zero frequency Fourier components of fields have
very short correlation lengths (of order $T^{-1}$) and may
be integrated out perturbatively.
Fermion fields, being antiperiodic, have no static components
and may be completely eliminated.
Once the fermions are gone, the SU(3) gauge field is decoupled
and may be dropped, as may the U(1) gauge field if the small
weak mixing angle is treated as perturbation.
This leaves an SU(2)-Higgs theory, which is straightforward
to define on a lattice.

    During the past two years, several groups have
performed large scale simulations of the finite temperature
phase transition in SU(2)-Higgs theory.
The authors of refs. \cite {Kripfganz,Fodor&Co} have chosen to
simulate a four-dimensional SU(2)-Higgs theory with a periodic ``time''
dimension consisting of $N_t = 2, 3, \ldots$ lattice spacings.
In contrast, the authors in refs. \cite {Kajantie,Farakos}
begin with the effective three dimensional theory which
results from integrating out all non-static Fourier components.
Spatial lattice sizes in these simulations are substantial
(typically $16^3$ to $32^3$).

    I will not attempt to summarize all the results of these
lattice simulations (which include data on the transition location
and character, latent heat, surface tension, and correlation lengths).
Interested readers should refer to the latest papers.
However, if one asks what lessons can be extracted from
the results of these simulations,
the the following points appear to be fairly well established:
\begin {itemize}\itemsep 4pt
\item [\it a.]
    Reasonably good calculations can be performed,
    at least for light Higgs masses.
    Finite lattice size and non-zero lattice spacing errors
    are (at least beginning to be) under reasonable control.
\item [\it b.]
    For Higgs masses below $\mw$,
    a first order phase transition is seen.
    The discontinuity at the transition becomes steadily
    weaker with increasing Higgs mass.
\item [\it c.]
    When $\mh \lt 20$ GeV, the lattice results agree quite well
    with perturbative predictions.
    When $\mh \gsim 50$ GeV, the strength of the transition
    appears to be larger than two-loop perturbative predictions.
\item [\it d.]
    Simulations
    (of sufficient accuracy to study a weakly discontinuous transition)
    become progressively more difficult as the Higgs mass grows.
    So far, no convincing conclusions can be drawn about the
    nature of the transition when $\mh \gsim \mw$.
\end {itemize}

\section {$\epsilon$-expansions}

    The $\epsilon$-expansion provides an alternative
systematic approach for computing the effects of (near)-critical
fluctuations.
It is based on the idea that
instead of trying to solve a theory directly in three spatial dimensions,
it can be useful to generalize the theory
from 3 to $4{-}\epsilon$ spatial dimensions,
solve the theory near four dimensions (when $\epsilon \ll 1$),
and then extrapolate to the physical case of 3 spatial dimensions.
Specifically, one expands physical quantities in powers of $\epsilon$
and then evaluates the resulting (truncated) series at $\epsilon = 1$
\cite {Wilson}.
This can provide a useful approximation when the
relevant long distance fluctuations are weakly coupled near 4 dimensions,
but become sufficiently strongly coupled that the loop expansion parameter
is no longer small in three dimensions.

    Scalar $\phi^4$ theory (or the Ising model) is a classic example.
In four dimensions, the long distance structure of a quartic scalar
field theory is trivial;
this is reflected in the fact that the renormalization group equation
$
    \mu (d\lambda / d\mu) = c \, \lambda^2 + O(\lambda^3) \,,
$
has a single fixed point at $\lambda = 0$.
In $4{-}\epsilon$ dimensions, the canonical dimension of the field changes
and the renormalization group equation acquires a linear term,
$$
    \mu {d\lambda \over d\mu} = -\epsilon \, \lambda + c \, \lambda^2 +
    O(\lambda^3) \,.
$$
This has a non-trivial fixed point
(to which the theory flows as $\mu$ decreases)
at $\lambda^* = \epsilon / c + O(\epsilon^2)$.
The fixed point coupling is $O(\epsilon)$ and thus small
near four dimensions, but grows with decreasing dimension and
becomes order one when $\epsilon = 1$.
Near four dimensions,
a perturbative calculation in powers of $\lambda$ is reliable
and directly generates an expansion in powers of $\epsilon$.

The existence of an infrared-stable fixed point indicates the
presence of a continuous phase transition as the
bare parameters of the theory are varied.
Performing conventional (dimensionally regularized) perturbative
calculations
and evaluating the resulting series at the fixed point,
one finds, for example, that the susceptibility exponent
(equivalent to the anomalous dimension of $\phi^2$) has the expansion
\cite {Wilson,Gorishny}
\begin {equation}
\label {Igamma}
   \gamma = 1 + 0.167 \, \epsilon + 0.077 \, \epsilon^2 - 0.049 \, \epsilon^3
   + O(\epsilon^4) \,.
\end {equation}
Adding the first three non-trivial terms in this series,
and evaluating at $\epsilon=1$,
yields a prediction which agrees with the best available result
to within a few percent
\cite {Zinn-Justin,Gupta}.

Inevitably, perturbative expansions in powers of $\lambda$
are only asymptotic;
coefficients grow like $n!$, so that succeeding terms in the series
begin growing in magnitude when
$
    n \gsim O(1/\lambda)
$.
Expansions in $\epsilon$ are therefore also asymptotic,
with terms growing in magnitude beyond some order
$n \gsim O(1/\epsilon)$.
If one is lucky, as is the case in the pure scalar theory,
$O(1/\epsilon)$ really means something like three or four
when $\epsilon = 1$ and the first few terms of the series will be useful.
If one is unlucky, no terms in the expansion will be useful.
Whether or not one will be lucky cannot be determined in advance of an
actual calculation.


    To apply the $\epsilon$-expansion to electroweak theory,
one begins with the full $3{+}1$ dimensional finite temperature
Euclidean quantum field theory (in which one dimension is periodic
with period $\beta = 1/T$) and integrates out all non-static Fourier
components of the fields.
The integration over modes with momenta of order $T$ or larger
may be reliably performed using standard perturbation theory in
the weakly-coupled electroweak theory.
This reduces the theory to an effective 3-dimensional SU(2)-Higgs theory.%
%
\footnote
    {%
    Fermions, having no static Fourier components, are completely
    eliminated in the effective theory.
    For simplicity, the effects of a non-zero weak mixing angle
    and the resulting perturbations due to the U(1) gauge field
    are ignored.
    Finally, one may also integrate out the static part of
    the time component of the gauge field, since this field
    acquires an $O(gT)$ Debye-screening mass.
    }
The effective theory depends on three relevant renormalized parameters:
\begin {center}
\begin {tabular}{rl}
    $g_1(T)^2_{\strut}$		\quad---& the SU(2) gauge coupling,\\
    $\lambda_1(T)_{\strut}$	\quad---& the quartic Higgs coupling,\\
    $m_1(T)^2$			\quad---& the Higgs mass (squared).
\end {tabular}
\end {center}

\begin {figure}
\vbox
    {%
    \begin {center}
	\leavevmode
	
	\epsfbox [72 260 520 550] {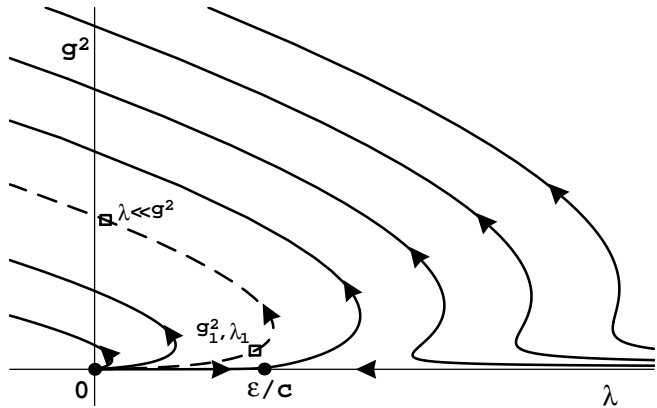}
    \end {center}
    \caption
	{%
	\label {figc}
	The renormalization group flow for an SU(2)-Higgs theory.
	Arrows indicate the direction of decreasing renormalization point.
	The dashed line is the trajectory which flows from an initial set
	of couplings $(g_1^2, \lambda_1)$ into the region where
	$\lambda \ll g^2$.
	}%
    }%
\end {figure}

    Next, one replaces the 3-dimensional theory by the corresponding
$4{-}\epsilon$ dimensional theory (and scales the couplings so that
$g_1^2 / \epsilon$ and $\lambda_1 / \epsilon$ are held fixed).
This is the starting point for the $\epsilon$-expansion.
When $\epsilon$ is small, one may reliably compute the
renormalization group flow of the effective couplings.
The renormalization group equations have the form
\begin {eqnarray}
    \mu {d\lambda \over d\mu}
&=&
    -\epsilon \, \lambda
    + (a \, g^4 + b \, g^2 \lambda + c \, \lambda^2) + \cdots \,,
\label {RG-a}
\\
    \mu {dg^2 \over d\mu}
&=&
    -\epsilon \, g^2
    + \beta_0 \, g^4 + \cdots \,.
\label {RG-b}
\end {eqnarray}
The precise values of the coefficients (and the next order terms)
may be found in reference \citenum {A&Y}.
These equations may be integrated analytically, and produce the
flow illustrated in figure \ref {figc}.

    Note that a non-zero gauge coupling renders the Ising fixed point
at $\lambda = \epsilon/c$ unstable, and that no other
(weakly coupled) stable renormalization group fixed point exists.
Trajectories with $g^2 \gt 0$ eventually cross the $\lambda = 0$ axis
and flow into the region where the theory (classically) would appear
to be unstable.
Such behavior is typically indicative of a first-order phase transition
\cite {Ginsparg}.
To determine whether this is really the case,
one must be able to perform a reliable calculation of the effective
potential (or other physical observables).
As discussed earlier,
the loop expansion parameter
for long distance physics is $\lambda(\mu) / g^2(\mu)$.
Consequently, the best strategy is to use the renormalization group
to flow from the original theory at $\mu = T$, which may have
$\lambda(T) / g^2(T)$ large, to an equivalent theory with $\mu \ll T$
for which $\lambda (\mu) / g^2(\mu)$ is small.
This is equivalent to the condition that one decrease the renormalization
point until it is comparable to the relevant scale for long distance physics,
specifically, the gauge boson mass, $M$.
By doing so, one eliminates large factors of
$\left[ (M / \mu)^\epsilon - 1 \right] / \epsilon$
which would otherwise spoil the reliability of the loop expansion.
(This, of course, is nothing other than the transcription to $4{-}\epsilon$
dimensions of the usual story in 4 dimensions, where
appropriate use of the renormalization group allows one to sum up
large logarithms which would otherwise spoil the perturbation expansion.)

    For small $\epsilon$, the change of scale required to flow from
an initial theory where $\lambda_1 / g_1^2 = O(1)$ to an equivalent theory
with $\lambda(\mu) / g^2(\mu) \ll 1$ is exponentially large;
the ratio of scales is
$$
    s \equiv {T \over \mu}
    \sim e^{\lambda_1 / g_1^4}
    \sim e^{O(1/\epsilon)} \,.
$$

    Given the parameters $g^2(\mu)$, $\lambda(\mu)$ and $m^2(\mu)$
of the resulting effective theory, one may use the usual loop expansion
to compute interesting physical quantities.
Because the change in scale is exponentially sensitive to $1/\epsilon$,
the result for a typical physical quantity will have the schematic form
\begin {eqnarray}
    {\cal O}
&=&
    f[g^2(\mu), \lambda(\mu), m^2(\mu)]
    \left( {\mu \over T} \right)^\#
\\
&\sim&
    \epsilon^\# \left( 1 + O(\epsilon) + \cdots \right) \;
    \exp \! \left[
	{\# \over \epsilon} \left( 1 + O(\epsilon) + \cdots \right)
    \right] \,.
\label {expansion}
\end {eqnarray}
In general, a calculation accurate to $O(\epsilon^n)$ requires
an $n$-loop calculation in the final effective theory, together
with $n{+}1$ loop renormalization group evolution.

To obtain predictions for the original theory in three spatial
dimensions, one finally truncates the expansions at a given order
and then extrapolates from $\epsilon \ll 1$ to $\epsilon = 1$.
Just as for the simple $\phi^4$ theory, the reliability of the resulting
predictions at $\epsilon = 1$ can only be tested {\em a-posteriori}.

This procedure has been carried out for
a variety of observables characterizing the electroweak
phase transition at both leading and next-to-leading order in the
$\epsilon$-expansion \cite {A&Y}.
Leading order results are available for the scalar correlation length
in both symmetric and asymmetric phases,
the free energy difference $\Delta F(T)$
between the symmetric and asymmetric phases,
the latent heat $\Delta Q = -T (d \Delta F / dT) |_{\Tc}$,
the surface tension $\sigma$ between symmetric and asymmetric phases at $\Tc$,
the bubble nucleation rate $\Gamma_{\rm N}(T)$ below $\Tc$,
and the baryon violation (or sphaleron) rate $\Gamma_{\Delta {\rm B}}(\Tc)$.

    The lowest order $\epsilon$-expansion predictions differ from
the results of standard one-loop perturbation theory (performed directly
in three space dimensions) in several interesting ways.
First, the $\epsilon$-expansion predicts a {\em stronger} first order
transition than does one-loop perturbation theory
(as long as $M_{\rm H} \lt 130$ GeV).
The correlation length at the transition is smaller, and the latent
heat larger, than the perturbation theory results.
The size of the difference depends on the value of the
Higgs mass (see reference \citenum {A&Y} for quantitative results).
Naively, one would expect that a stronger first order transition
would imply a smaller baryon violation rate, since a larger
effective potential barrier between the co-existing phases should
decrease the likelyhood of thermally-activated transitions
across the barrier.
This expectation is wrong (in essence, because it unjustifiably
assumes that the shape of the barrier remains unchanged).
Along with predicting a strengthing of the transition,
the $\epsilon$-expansion predicts a {\em larger}
baryon violation rate.
This occurs because the baryon violation rate is exponentially
sensitive to the sphaleron action (or mass),
$$
    \Gamma_{\Delta {\rm B}} \propto \exp -S_{\rm sphaleron} \,,
$$
and the sphaleron action depends inversely on $\epsilon$,
$$
    S_{\rm sphaleron} = {\# \over g^2(\mu)} = O(1/\epsilon) \,.
$$
Hence, unlike other observables, the exponential sensitivity to
$1/\epsilon$ in the baryon violation rate does not arise solely
from an overall power of the scale factor $\mu / T$.

Note that an increase in the baryon violation rate (compared to
standard perturbation theory) makes the constraints
for viable electroweak baryogenesis more stringent;
specifically, the (lowest order) $\epsilon$-expansion suggests that
the minimal standard model bound $M_{\rm H} \lsim 35$--40
GeV derived using one-loop perturbation
theory in ref.~\citenum {Dine} should be even lower,
further ruling out electroweak baryogenesis in the minimal model.


    As emphasized earlier, in general there is no way to know,
in advance of an actual calculation, how many terms (if any)
in an $\epsilon$-expansion will be useful when results are
extrapolated to $\epsilon = 1$.
Therefore, to assess the reliability of an $\epsilon$-expansion
one must be able to test predictions for actual physical quantities.
For the electroweak theory, two types of tests are possible:
\begin {itemize}
\item [\bf A.]\itemsep 10pt
    $\lambda \ll g^2$.
    In the limit of a light (zero temperature) Higgs mass,
    or equivalently small $\lambda_1 / g_1^2$,
    the loop expansion in three dimensions is reliable.
    Hence, although this is not a realistic domain, one may easily
    test the reliability of the $\epsilon$-expansion in this regime by
    comparing with direct three-dimensional perturbative calculations.
    Table~1 summarizes the fractional error for various
    physical quantities produced by truncating the $\epsilon$-expansion
    at leading, or next-to-leading, order before evaluating at
    $\epsilon = 1$, in the light Higgs limit.
    Although the lowest-order results often error by a factor of two
    or more, all but one of the next-to-leading order results
    are correct to better than 10\%.
    (The free energy difference at the limit of metastability,
    $\Delta F(T_0)$, has the most poorly behaved $\epsilon$-expansion.
    However, if one instead computes the logarithm of this quantity,
    then the next-to-leading order result is correct to within 17\%.
    The baryon violation rate is not shown because, due to the way
    its $\epsilon$-expansion was constructed, the result is trivially
    the same as the three-dimensional answer when $\lambda_1 \ll g_1^2$.
    See ref.~\citenum {A&Y} for details.)
    \begin{table}[h]
    \begin {center}
    \def\myline{\noalign{\vskip-3pt}\hline\noalign{\vskip-3pt}}
    \def\stick{\vphantom{\Big|}}
    \def\stickup{\hbox{\vrule height 11pt depth 0pt width 0pt}}
    \def\stickdown{\hbox{\vrule height 0pt depth 5pt width 0pt}}
    \tabcolsep=8pt
    \begin {tabular}{|lc|cc|}             \myline
    \multicolumn{1}{|c}{observable ratio\stick}
      &                   & LO          & NLO         \\ \myline
    asymmetric correlation length\stickup
      &   $\xi\asym$      & \phantom-0.14        & -0.06        \\
    symmetric correlation length
      &   $\xi\sym$       & \phantom-0.62        & -0.08        \\
    latent heat
      &   $\Delta Q$      & -0.23        	&\phantom-0.04	\\
    surface tension
      &   $\sigma$        & -0.40        	& -0.02        \\
    free energy difference\stickdown
      &   $\Delta F(T_0)$ & -0.76        	& -0.44        \\ \myline
    \end {tabular}
    \end {center}
    \caption
	{%
	\label {tableb}
	The fractional error in the $\eps$-expansion results,
	when computing prefactors through leading order (LO) and
	next-to-leading order (NLO) in $\eps$,
	when $\lambda_1 \ll g_1^2$.
	}%
    \end{table}
\item [\bf B.]
    $\lambda \gsim g^2$.
    When $\lambda / g^2$ is $O(1)$, the three-dimensional loop expansion
    is no longer trustworthy.
    However, one may still test the stability of $\epsilon$-expansion
    predictions by comparing $O(\epsilon^n)$ and $O(\epsilon^{n+1})$
    predictions --- provided, of course, one can evaluate at least two
    non-trivial orders in the $\epsilon$-expansion.
    For most physical quantities this is not (yet) possible;
    determining the lowest-order behavior of the prefactor in
    expansion (\ref {expansion}) requires a one-loop calculation
    in the final effective theory together with a two-loop evaluation
    of the (solution to the) renormalization group equations.
    A consistent next-to-leading order calculation requires a two-loop
    calculation in the final theory together with three-loop
    renormalization group evolution.
    Althouth two-loop results for the effective potential and
    beta functions are known, three loop renormalization group
    coefficients in the scalar sector are not currently available.
    Nevertheless, by taking suitable combinations of physical quantities
    one can cancel the leading dependence on the scale ratio
    $\mu / T$ and thereby eliminate the dependence
    (at next-to-leading order) on the three loop beta functions.
    For example, the latent heat depends on the scale as
    $\Delta Q \sim (\mu / T)^{2+\epsilon}$ while the
    scalar correlation length $\xi \sim (\mu / T)^{-1}$.
    Therefore, the combination $\xi^2 \Delta Q$ cancels the leading
    $\mu / T \sim e^{O(1/\epsilon)}$ dependence and thus requires
    only two-loop information for its next-to-leading order evaluation.
    The result of the (rather tedious) calculation may be put in
    the form
    \begin {equation}
	\xi\asym^2 \Delta Q
	=
	T^{1-\epsilon} \, f(f_1^2, \lambda_1)
	\left[ 1 + \delta + O(\epsilon^2) \right] \,,
    \end {equation}
    where $\delta$, the relative size of the next-to-leading order
    correction, is plotted in figure \ref {figlat}.
    The correction varies between roughly $\pm30\%$ for
    (zero temperature) Higgs masses up to 150 GeV.
    This suggests that the $\epsilon$ expansion is tolerably
    well behaved for these masses.
    For larger masses the correction does not grow indefinitely,
    but is bounded by 80\%,
    suggesting that the $\epsilon$ expansion may remain
    qualitatively useful even when it does not work as well
    quantitatively.
    \begin {figure}[h]
    \vbox
	{%
	\begin {center}
	    \leavevmode
	    
	    \epsfbox [150 250 500 500] {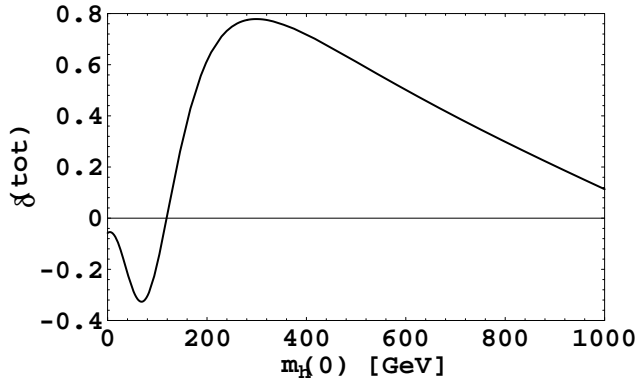}
	\end {center}
	\caption
	    {%
	    \label {figlat}
	    The relative size of the next-to-leading order correction
	    to $\xi\asym^2 \Delta Q$ in the $\eps$-expansion.  The values
	    are given as a function of the (tree-level)
	    zero-temperature Higgs mass in minimal SU(2)
	    theory ($N=2$) with $g = 0.63$.
	    }%
	}%
    \vspace* {-.1in}
    \end {figure}

    \end {itemize}

\section {Conclusions}

The most useful technique for studying the electroweak phase transition
clearly depends on the range of value of the Higgs mass.
For sufficiently light Higgs ($\mh \lsim 40$ GeV)
perturbation theory is adequate.
At intermediate masses ($40\;{\rm GeV} \lsim \mh \lsim 80$ GeV)
numerical simulations have been quite effective.
For heavier Higgs masses, the $\epsilon$-expansion appears promising.

Although considerable progress has been made toward a quantitative
understanding of the electroweak phase transition, much remains to be done.
With continuing efforts, numerical results will undoubtedly improve,
particularly for Higgs masses above $\mw$.
Calculations of additional physical quantities at next-to-leading
order in the $\epsilon$-expansion should definitely be performed
to further confirm the reliability of the method.

Nevertheless, it should be noted that quantitative understanding of the
phase transition is already reasonably good in the range of Higgs masses
for which the transition is sufficiently discontinuous to be
compatible with baryogenesis.
Hence,
the major source of uncertainty about the viability of
electroweak baryogenesis
appears to be the lack of knowledge about which theory is correct
(extra Higgs, supersymmetry, {\em etc}.)
not the limitations in our ability to compute
the thermodynamics of these theories.

\end {document}